\documentclass[fleqn,usenatbib]{mnras}

\usepackage{newtxtext,newtxmath}
\usepackage[T1]{fontenc}
\usepackage{ae,aecompl}

\usepackage{graphicx}	% Including figure files
\usepackage{amsmath}	% Advanced maths commands

\usepackage{amssymb}	% Extra maths symbols
\usepackage{xcolor}
\usepackage{hyperref}
\usepackage{orcidlink}

\newcommand{\orcid}[1]{\href{https://orcid.org/#1}{\textcolor[HTML]{A6CE39}{\aiOrcid}}}

\def\la{\mathrel{\mathpalette\fun <}}
\def\ga{\mathrel{\mathpalette\fun >}}
\def\fun#1#2{\lower3.6pt\vbox{\baselineskip0pt\lineskip.9pt
        \ialign{$\mathsurround=0pt#1\hfill##\hfil$\crcr#2\crcr\sim\crcr}}}

\title[Roman galaxy redshift survey] {Simulating properties of emission line galaxies from Nancy Grace Roman Space Telescope}

\author[Z. Zhai et al.]{\parbox{\textwidth}{
Zhongxu Zhai,$^{1}$\thanks{E-mail: zhai@ipac.caltech.edu}
Yun Wang,$^{1}$
Andrew Benson,$^{2}$
James Colbert,$^1$
Micaela Bagley, $^{3}$ \orcidlink{0000-0002-9921-9218}
Alaina Henry, $^{4}$
Ivano Baronchelli, $^{5}$ 
}\vspace*{4pt}\\
$^{1}$IPAC, California Institute of Technology, Mail Code 314-6, 1200 E. California Blvd., Pasadena, CA 91125 \\
$^{2}$Carnegie Observatories, 813 Santa Barbara Street, Pasadena, CA 91101 \\
$^{3}$Department of Astronomy, The University of Texas at Austin, Austin, TX, USA \\
$^{4}$Space Telescope Science Institute; 3700 San Martin Drive, Baltimore, MD, 21218  \\
$^{5}$INAF - Osservatorio Astronomico di Padova, Vicolo dell'Osservatorio 5, I-35122, Padova, Italy
}
\date{Accepted XXX. Received YYY; in original form ZZZ}

\pubyear{2020}

\begin{document}
\label{firstpage}
\pagerange{\pageref{firstpage}--\pageref{lastpage}}
\maketitle

\begin{abstract}

We simulate the emission line galaxy properties for the Nancy Grace Roman Space Telescope by creating a 4 deg$^{2}$ galaxy mock catalog using Galacticus, a semi-analytic model (SAM) for galaxy formation and evolution. The simulated galaxy properties include emission line luminosity, equivalent width (EW), broad band photometry, and spectral energy distribution (SED). We compare this mock catalog with the latest observational data from WISP and MOSDEF.
We find that our Galacticus model makes predictions consistent with observational data, over a wide redshift range for multiple emission lines. 
We use this validated galaxy mock catalog to forecast the photometric completeness of H$\alpha$ and [OIII] emission lines as a function of line flux cut and H band magnitdue cut, for both Roman and Euclid-like surveys. Our prediction for the photometric completeness of a Euclid-like survey is in good agreement with previous work based on WISP data. Our predictions for the photometric completeness of possible implementations of the Roman High Latitude Wide Area Spectroscopic Survey (HLWASS) provides key input for the planning of such a survey. We find that at H = 24, a Euclid-like survey to the line flux limit of 2$\times 10^{-16}\,$erg/s/cm$^2$ is 97\% complete, while a Roman HLWASS to the line flux limit of $10^{-16}\,$erg/s/cm$^2$ is only 94.6\% complete (it becomes 98\% complete at H = 25).

\end{abstract}


\section{Introduction}

Galaxy redshift surveys play an important role in cosmology. The mapping of the spatial distribution of a huge number of galaxies over cosmic volumes reveals large coherent structures evolved under gravitational interaction. Investigations of their clustering properties enable various cosmological measurements at different epochs, including the cosmic distance scales and expansion history through baryon acoustic oscillation (BAO), and growth history of structure through redshift space distortion (RSD) effect. These measurements provide important information to constrain the viable parameter space of the standard $\Lambda$CDM model and any deviations from it, including numerous models of dark energy and modified gravity theories which were proposed to explain the observed cosmic acceleration (\citealt{Riess_1998, Perlmutter_1999}). In order to better understand the mysteries, many ongoing and future survey missions were proposed such as Dark Energy Spectroscopic Survey (DESI, \citealt{DESI_2016}), Euclid (\citealt{Laureijs_2011, Laureijs_2012}), and NASA's Nancy Grace Roman Space Telescope (hereafter Roman, \citealt{Green_2012, Dressler_2012, Spergel_2015}). These programs will significantly improve the accuracy of cosmological measurements and explore the expected constraints on the properties of dark energy. 

As NASA's flagship survey for next generation, Roman will measure millions of emission line galaxies (ELG) as tracers of underlying matter distribution. In particular, Roman will observe H$\alpha$ and [OIII] emitters as the main targets within the redshift range $1.0<z<3.0$. Its design will be complementary to that of Euclid (shallow and wide over 15,000 deg$^2$), by going deeper over a smaller survey area. In addition to the three-dimensional mapping of the large scale structure, Roman can also measure the star formation history of millions of galaxies to reveal the physics for galaxy formation and evolution. The measurement of large scale structure and the implications for cosmology from Roman can be realized through a large area slitless spectroscopic survey at high Galactic latitude, the High Latitude Wide Area Spectroscopic Survey (HLWASS). A description of a worked example of the HLWASS, the High Latitude Spectroscopic Survey (HLSS), is presented in \cite{Wang_2021} and we refer the reader to that paper for more details. 

In order to maximize the science return of the space mission, it is necessary to realistically forecast its capability and optimize the survey strategies. \cite{Bagley_2020} construct a sample using WISP survey and forecast the properties of H$\alpha$ and [OIII] emitting galaxies from future surveys. \cite{Bowman_2021} adopt a similar method based on the data from HST/WFC3 G141 grism and measure the luminosity function of [OIII] galaxies for both Euclid and Roman. \cite{Baronchelli_2020} apply a machine learning algorithm on the identification of single emission line which can improve the survey completeness for spectroscopic surveys. Earlier work based on the earlier observational data from WISP can be found in \cite{Colbert_2013, Mehta_2015}. In addition to the forecast using available observational data, it is also useful to employ numerical simulations to produce realistic mock catalogs. In \cite{Merson_2018}, the authors apply Galacticus (\citealt{Benson_2012}), a semi-analytical model (SAM) of galaxy formation to produce a synthetic galaxy catalog and forecast the number densities of H$\alpha$ emitters with different dust attenuation models. In \cite{Zhai_2019MNRAS}, we recalibrate the SAM using the state of the art N-body simulation suite, UNIT (\citealt{Chuang_2019}) and make a 4 deg$^2$ mock for Roman HLSS, and forecast the number densities of H$\alpha$ and [OIII] emitters as a function of redshift. Using the same SAM, we produce a larger galaxy mock covering 2000 deg$^2$ sky area and investigate the clustering properties of H$\alpha$ galaxies (\citealt{Zhai_2020}); we analyze this mock by applying methods used in analyzing real data, and evaluate the significance of BAO and RSD measurements. This cosmology mock is further applied in \cite{Zhai_2021}, we measure the linear bias of H$\alpha$ and [OIII] galaxies for Roman HLSS with different sample selections. This can provide critical information for a following Fisher matrix analysis to forecast the ability of the survey to constrain cosmology and dark energy (\citealt{Wang_2008b}).

The SAM we have used is calibrated to match the observed H$\alpha$ luminosity function from the ground-based narrow-band High-z Emission Line Survey (HiZELS, \citealt{Geach_2008, Sobral_2009, Sobral_2013}). Our current application of this model focuses on the emission line luminosity or flux of H$\alpha$ and [OIII]. However, we should note that this model also needs comprehensive comparison with available observational data, such as the broad band magnitude, equivalent width (EW) of emission lines, distribution of color-color relation and so on. This can verify if the SAM can make robust and self-consistent predictions of the galaxy properties. Similar to other SAM, the Galacticus model we have used couples the physical processes of star formation, gas cooling, AGN feedback, galaxy mergers and so on. Embedding these processes into the background of a high-resolution N-body simulation for dark matter can give detailed properties of galaxies, which enables the comparison of the detailed galaxy properties. 
In this paper, we apply the Galacticus model to make such a comparison. This model is also coupled with stellar population spectral templates to output spectral energy distribution (SED) within the wavelength range of the Roman grism, as the input for realistic grism simulation. By combining it with observational effects such as exposure time, detection limit, roll angles and dithering, the resultant product of the Roman grism simulation can be as close as possible to actual future observations and can be used to characterize the observational systematics. We will leave this for future work. 

In addition to the examination of the SAM, we should note that these observational data can be used to calibrate the model parameters. Since our primary goal of the SAM calibration is to forecast the number density of ELGs from Roman HLSS, only the relevant H$\alpha$ luminosity function is used. As more observational data become available, adding them into the calibration process can improve the overall performance of the model, and reveal any correlation or tension between the physical processes of galaxy formation.

For future galaxy surveys like Roman HLSS and Euclid, one of the critical features is the completeness of the galaxy sample under certain flux limit. Since the spectral extraction of galaxies is only performed above a certain continuum magnitude, some galaxies can be missed even though they have bright enough emission lines. This has a direct impact on the galaxy density and shot noise of the statistics. Based on the constructed catalog from WISP, \cite{Bagley_2020} estimate the completeness of H$\alpha$+[NII] and [OIII] for a Euclid-like survey and show that the fraction of missing galaxies is 2\% for $H=24$. With the detailed galaxy properties from our SAM catalog, we can investigate the completeness of Roman HLSS under different observing conditions, which can provide additional information for the survey design and strategies. Incorporation of this estimate can make the following forecast for dark energy constraint in a more realistic manner. 

Our paper is organized as follows. \S2 introduces the Roman HLSS, a worked example of the Roman High Latitude Wide Area Spectroscopic Survey (HLWASS). \S3 gives an overview of Galacticus, a state-of-the-art semi-analytical model for galaxy formation and evolution, and describes its calibration using available observational data. \S4 describes the observational data used to compare with Galacticus predictions. \S5 presents our results on the ELG properties for Roman predicted by Galacticus, and how they compared with observational data. \S6 presents our results on the photometric completeness for Roman, as a function of emission line flux limit. We summarize and conclude in \S7.

\section{Roman HLSS}

We start with a brief introduction of Roman HLSS in this section, a detailed description can be found in \cite{Wang_2021}. The Roman HLSS, the only fully worked example of the Roman High Latitude Wide Spectroscopic Survey; the actual HLWASS that Roman will execute will be determined by NASA in an open community process. 

As a key component of the Roman observing program, HLSS has the main goal of answering the question of dark energy that causes the cosmic acceleration. The primary measurements are the expansion history of the universe and the growth history of large scale structure. The results can also serve as cross-checks with other analysis techniques, such as supernova, weak lensing and galaxy clusters. A joint analysis can further tighten the constraint on the fundamental parameters of cosmology. 

During its 6 months of observations, Roman HLSS will survey $\sim$ 2000 deg$^2$ area. This will be sufficient for a robust measurement of BAO and RSD signals over the entire redshift range. More than 90\% of the survey area will also be observed by the  High Latitude Imaging Survey (HLIS). This will not only maximize the science return of dark energy research via a joint analysis of galaxy clustering and weak lensing, but also improve the quality of the data with better redshift determination.

The HLSS grism has a wavelength coverage of $1-1.93 \mu$m, which enables the detection of both H$\alpha$ (656.3nm) and [OIII] (500.7nm) at $1.1<z<1.9$ to assure a robust redshift determination. At higher redshift where H$\alpha$ is outside the window, the emission lines [OIII] (500.7nm) and [OII] (372.6nm) can be used for redshift measurement; this  extends the Roman redshift coverage to $z\sim 2.8$. Note that the Euclid grism has the red cutoff at $1.85\mu$m, bluer than the Roman grism cutoff at $1.93 \mu$m, thus Roman has greater redshift reach by design. In addition, the shallower depth of Euclid only allows robust detection of H$\alpha$ up to redshift $0.9<z<1.8$, with insignificant numbers of [OIII] at higher redshifts. Therefore the Roman HLSS is complementary to that of Euclid in terms of both sensitivity and redshift coverage.

Based on our previous forecast work using Galacticus (\citealt{Zhai_2019MNRAS}), for a flux limit of $1.0\times10^{-16}\mathrm{erg}~\mathrm{s}^{-1}\mathrm{cm}^{-2}$ at 6.5$\sigma$, Roman HLSS can observe more than 10M H$\alpha$ ELGs with $1<z<2$, and another 2M [OIII] ELGs with $2<z<3$. The Roman science requirement of detecting both lines ensures a high purity sample for BAO/RSD measurements (\citealt{Zhai_2021}). Compared with the Euclid survey which has the line flux limit of $2.0\times10^{-16}\mathrm{erg}~\mathrm{s}^{-1}\mathrm{cm}^{-2}$ at $3.5\sigma$, Roman HLSS is designed to enable robust modeling of systematic effects.

The error of the redshift measurement is expected to have $\sigma_{z}<0.001(1+z)$ for Roman HLSS, excluding redshift outliers and galaxies with a radius greater than $0.54^{\prime\prime}$ (larger galaxies have larger redshift errors). Based on the WISP data, 90\% of the ELGs can satisfy this condition for both H$\alpha$ and [OIII] emissions. This overall requirement can suppress the systematic error of the subsequent BAO/RSD analysis to an acceptable level. In addition, the survey also requires the completeness higher than 60\%, with completeness defined as the fraction of galaxies with reliable redshifts which constitute the final sample for BAO/RSD measurements. This is based on extrapolation from Euclid, but needs validation using grism simulations. Our current mock can serve as input for this test and the resultant products will be presented in a future paper. Roman HLSS will include deep field observations over $\sim 22$ deg$^2$, for calibrating the completeness and purity of the wide field observations \citep{Wang_2021}. 

In addition to the default selection on emission line flux by construction via the grism slitless spectroscopy, Roman grism data pipeline will extract spectra for galaxies with continuum magnitude (e.g. $H$-band magnitude) above a certain limit, in order to ensure that appropriate NIR photometry from Roman is available for constructing the decontamination model for the slitless galaxy spectra. This means that not all the galaxies that can pass the emission line threshold can be observed. In this paper, we use Galacticus to investigate the photometric completeness of the Roman HLSS, defined as the ratio of the number of galaxies above a given continuum magnitude at a given line flux limit, over the total number of galaxies with emission lines above that line flux limit. Our results can inform the modeling of observational systematics, and guide the optimal data processing for the Roman HLSS.

\section{Galacticus model}

\begin{figure*}
\begin{center}
\includegraphics[width=17cm]{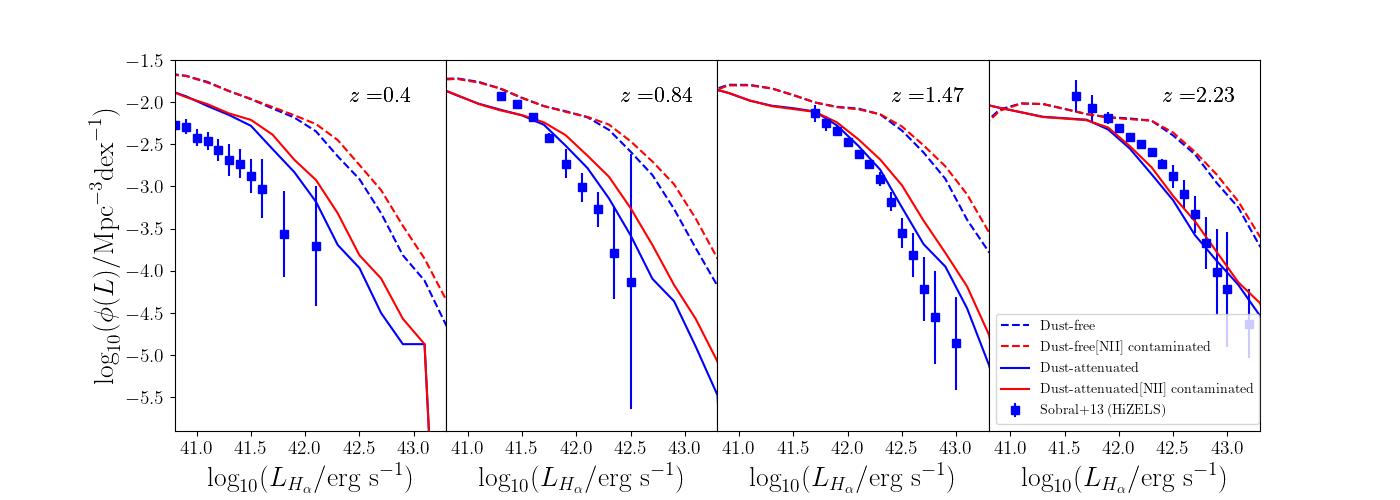}
\includegraphics[width=17cm]{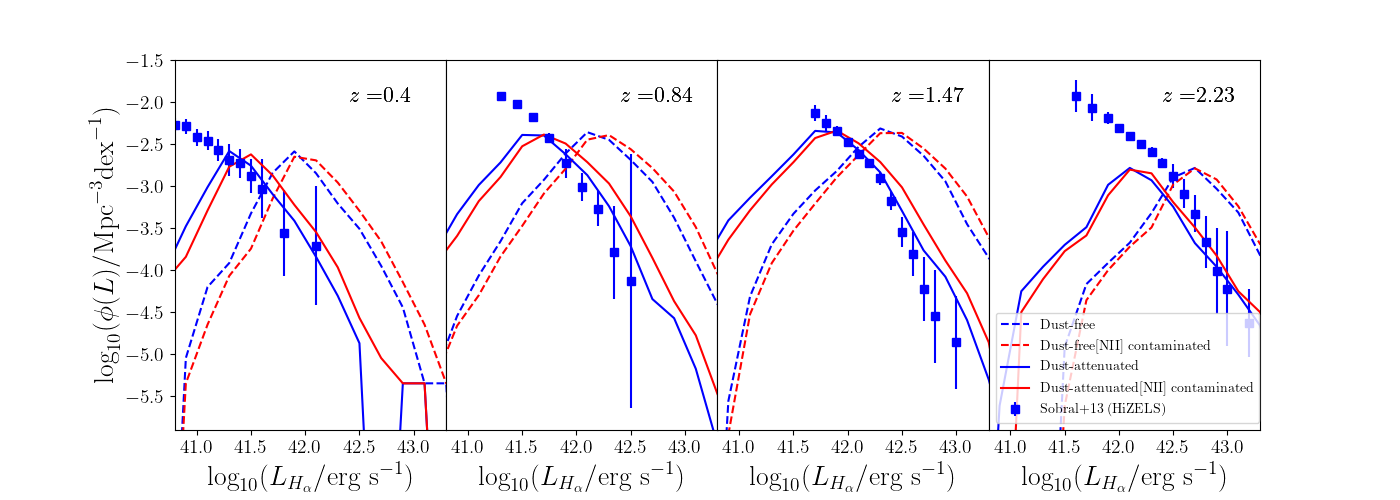}
\caption{The luminosity function of H$\alpha$ galaxies predicted from Galacticus. The top panel shows the raw prediction, while the bottom panel shows the result with selections applied as explained in the text. The dashed and solid lines represent dust-free and dust attenuated results respectively. The blue lines are H$\alpha$ only, and the red lines are [NII] contaminated.}
\label{fig:Halpha_LF}
\end{center}
\end{figure*}

The analysis of this paper is based on a synthetic galaxy catalog produced by Galacticus (\citealt{Benson_2012})--- a semi-analytical model (SAM) for galaxy formation and evolution. Compared with our previous works using the same methodology in \cite{Zhai_2019MNRAS, Zhai_2020, Zhai_2021}, the Galacticus software has been updated with better efficiency of ODE solver, and better control of numerical performance. These changes can affect some particular galaxy properties, however the galaxy statistics are only mildly influenced as we show below.

Similar to other SAM's, Galacticus works by parameterizing astrophysical processes governing galaxy formation and evolution, and performs their evolution in the background of a distribution of dark matter halos. Compared with hydro-dynamic simulation to model both dark matter and gas particles, Galacticus can produce a galaxy mock with a sufficiently large cosmic volume in a timely manner. The output can also have detailed galaxy properties such as the emission line luminosity, star formation history, galaxy morphology, spectral energy distribution and so on. The galaxy population can also be coupled with a set of filter transmission curves to obtain photometric luminosities and enable a direct comparison with observations. 

Calibration of the model parameters was performed in \cite{Zhai_2019MNRAS}, including the parameters of galaxy formation and the dust model of observed luminosities for both emission lines and continuum. In particular, the dust models were chosen to match the observed luminosity function of H$\alpha$ from the ground-based narrow-band High-z Emission Line Survey (HiZELS, \citealt{Geach_2008, Sobral_2009, Sobral_2013}), or the number counts data collected from Wide Field Camera 3 (WFC3) Infrared Spectroscopic Parallels survey (WISP; \citealt{Atek_2010, Atek_2011, Mehta_2015}). We use parameter $A_{V}$ of the \citet{Calzetti_2000} model to describe the strength of dust attenuation. For simplicity, we only present the WISP-based result with $A_{V}=1.7$ in this paper due to relatively larger sample size. The other HiZELS-based model $A_{V}=1.9$ gives slightly stronger dust attenuation. However, we believe that the results in the comparison with observational data between these dust models are not significantly different.

The Galacticus model takes the merger trees of dark matter halos from the UNIT simulation \footnote{\url{https://unitsims.ft.uam.es}} (\citealt{Chuang_2019}) as input. This suite assumes a spatially flat $\Lambda$CDM model with Planck 2016 cosmological parameters (\citealt{Planck_2016}). In particular, we choose the 1$h^{-1}$Gpc box for our work, which has a mass resolution of $10^{9}h^{-1}M_{\odot}$. The simulation product covers a redshift range of $0<z<99$, implying the SAM simulation for galaxies can be extended to higher redshift than Roman HLSS is designed to probe. We apply Galacticus to the merger trees extracted from this simulation and construct a 4 deg$^{2}$ light cone catalog of galaxies. We first compare the luminosity function of H$\alpha$ ELGs with the HiZELS observation, as shown in the top panel of Figure \ref{fig:Halpha_LF} to verify the calibration. The result shows that the model prediction has consistent amplitude over redshift range of $0.84<z<2.23$, which can fully cover the window of Roman HLSS. Although the HiZELS observations of Halpha emitters at z=0.4 fall a factor of 2 to 3 times below our model predictions, we note that the earlier luminosity function of \cite{Ly_2007} which is based on a much larger data sample, is $2\sim3$ times higher, and thus agrees well with our models. On the other hand, this discrepancy can be improved by applying a redshift dependent dust model with additional parameters, but can make the model more complex. Therefore we just proceed with the earlier model of constant dust parameter.

Roman HLSS has a spectral resolution high enough to separate the [NII] contamination from H$\alpha$ for most galaxies. However, this is not true for Euclid-like survey where H$\alpha$ and [NII] are blended for most galaxies. In order to explicitly show the impact of this on the luminosity function, we model the [NII] contamination using the empirical model from \cite{Faisst_2018} \footnote{The Galacticus model can also output [NII] emission line, however the current model prediction is lower than observation, implying a further improvement is needed. We thus leave the investigation of the [NII] emission to future work.}. This model introduces a redshift and stellar mass dependent correction due to [NII] emission based on SDSS galaxies. The result of this H$\alpha$+[NII] luminosity function is shown as the red line in the top panel of Figure \ref{fig:Halpha_LF}. Due to the boosted flux, the result is significantly higher than H$\alpha$ only, consistent with expectation. For a typical range of galaxies at $z=1.5$, this [NII] contamination can increase the luminosity function by up to 0.2 or 0.3 dex, and therefore significantly boosting the number of observed galaxies, see e.g. \cite{Merson_2018}.

\section{Observational data}

In this section, we introduce the observational data sets for the comparison with our Galacticus model predictions. We focus on the nebular emission lines and broad band magnitude at near-infrared, but note that the access to other observed galaxy properties can make more comparisons possible.

\subsection{WISP}

\begin{figure*}
\begin{center}
\includegraphics[width=16cm]{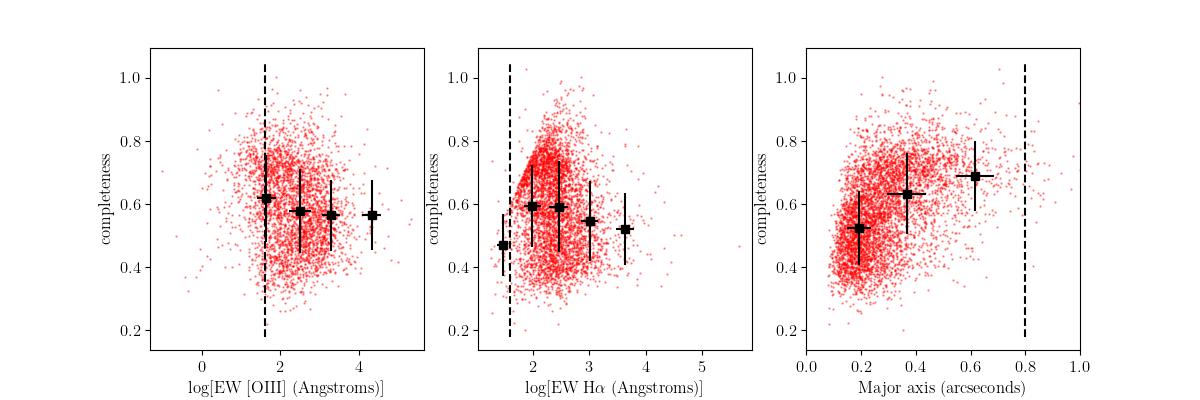}
\caption{Completeness of the emission line galaxies from the WISP catalog, as a function of [OIII] equivalent width (left), H$\alpha$ equivalent width (middle), and galaxy size (right). The red dots represent individual galaxies in the WISP catalog, then we bin the sample by [OIII] EW (left), H$\alpha$ EW (middle) and galaxy size (right), and compute the mean and standard deviation in each bin. The resultant measurements are shown as the black squares with errorbars. The dashed line represent the cuts we use for the following comparison.}
\label{fig:WISP_completeness}
\end{center}
\end{figure*}

\begin{figure}
\begin{center}
\includegraphics[width=8cm]{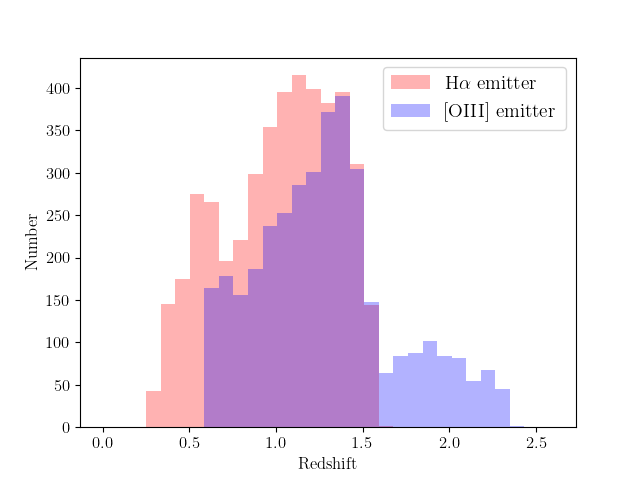}
\caption{Redshift distribution for H$\alpha$ emitters (red) and [OIII] emitters (blue. The overlap area can approximately represent the galaxies with both lines detected.}
\label{fig:z_distribution}
\end{center}
\end{figure}

The first data set we use is the emission line catalog from the WFC3 Infrared Spectroscopic Parallel survey (WISP,  \citealt{Atek_2010, Atek_2011, Mehta_2015}), which uses WFC3’s G102 and G141 grisms (0.8-1.7 um) to take slitless spectroscopy of a large number of random extragalactic fields, operating WFC3 in parallel to the other instruments aboard HST. The data sample we use includes emission-line measurements from 419 WISP fields and the sky coverage is $\sim1520$ arcmin$^{2}$. The construction of the WISP emission line catalog is described in detail by \cite{Bagley_2020}, but is based on an automatic detection algorithm combined with a visual inspection by two reviewers for each individual object in order to reject artifacts such as cosmic rays, hot pixels, and lines that are too heavily contaminated by overlapping spectra. The reviewers then identify the emission lines and fit the source redshift with a full spectrum including both emission lines and the continuum. The simultaneous fit of all emission lines ensures that all the emission line profiles are fit with the same FWHM, appropriate for slitless spectra where all emission lines are images of the same host source. The simultaneous fit also helps eliminate contamination from overlapping spectra. The creation of this WISP emission line catalog differs from earlier WISP catalogs (e.g., \citealt{Colbert_2013}) with its use of continuous wavelet transforms to fit not only the amplitude but also the shape of emission-line features in a spectrum. The new WISP emission line detection algorithm also includes additional quality checks to remove most of spurious sources before the inspection stage, although this mainly effects the total time required for visual inspection, rather than the end catalog product. The finalized WISP catalog includes $\sim8000$ emission line objects. 

In our analysis, we choose the cleanest WISP sample by selecting \textsl{redshift\_flag} < 2 as a conservative choice. This includes all objects with secure redshift determinations, i.e. both reviewers agree on the redshift, or if not, one of the reviewers identified three or more high S/N (>3) emission lines. This cut also excludes objects with redshifts determined by a single emission line. This selection reduces the sample size to 4324 objects. The process of the emission line construction and completeness analysis is presented in \cite{Bagley_2020}. In Figure \ref{fig:WISP_completeness}, we show the scatter distribution of the completeness as a function of [OIII], H$\alpha$ EW and galaxy size for the selected WISP catalog. This is analogous to Figure 1 of  \cite{Colbert_2013}, showing the dependence of recovery rate using simulation on the galaxy properties.

In Figure \ref{fig:z_distribution}, we present the redshift distribution of the H$\alpha$ and [OIII] emitters in the cleanest WISP catalog. The overlap area roughly represents galaxies that have both lines detected. The observation of reliable H$\alpha$ emitters can reach redshift up to $z\sim1.6$, while the [OIII] can extends to $z=2.4$. However there is a rapid decline at $z\sim1.6$ for the [OIII] emitters due to requirement of redshift determination since the primary emission H$\alpha$ is outside the HST grism window.

\subsection{MOSDEF}

The second data set is the emission line galaxies from the MOSFIRE Deep Evolution Field (MOSDEF) survey, compiled by \cite{Reddy_2018}. This catalog contains 1134 star-forming galaxies at redshift $1.4<z<3.8$ and enables the investigation of the relationships between the emission line equivalent width and other galaxy properties, such as stellar mass, age, star formation rate (SFR) and so on. The data compilation selects galaxies in three redshift windows: $1.37-1.70$, $2.09-2.70$ and $2.95-3.8$. The first two are consistent with the Roman observing window and therefore can provide a direct comparison. 

The survey selects objects from the 3D-HST photometric catalog with a limit of $H=24$, 24.5 and 25 for the above three redshift ranges. To be consistent with \cite{Reddy_2018}, we refer to these three subsamples as low-redshift, middle-redshift and high-redshift respectively in the following analysis. After correcting for the slit loss, emission line absorption, excluding object with low $S/N$ of the integrated line flux, only objects with secure spectroscopic redshifts from the MOSFIRE spectra are included in the final sample. This forms a largely representative sample of "typical" star-forming galaxies at redshift $1<z<4$ (\citealt{Reddy_2018}), which is also complete for line luminosities with stellar mass above $10^{9}M_{\odot}$. The galaxy properties including stellar mass, age and star formation rate are derived using photometric modeling, which allows the investigation of their correlations with emission line EW. For more details of the MOSDEF catalog including target selection, correction of observational effects and derivation of galaxy properties, we refer the readers to \cite{Reddy_2018}.

%\subsection{COSMOS}

\section{Results of comparison}

\subsection{Comparison with WISP}

\begin{figure*}
\begin{center}
\includegraphics[width=8cm]{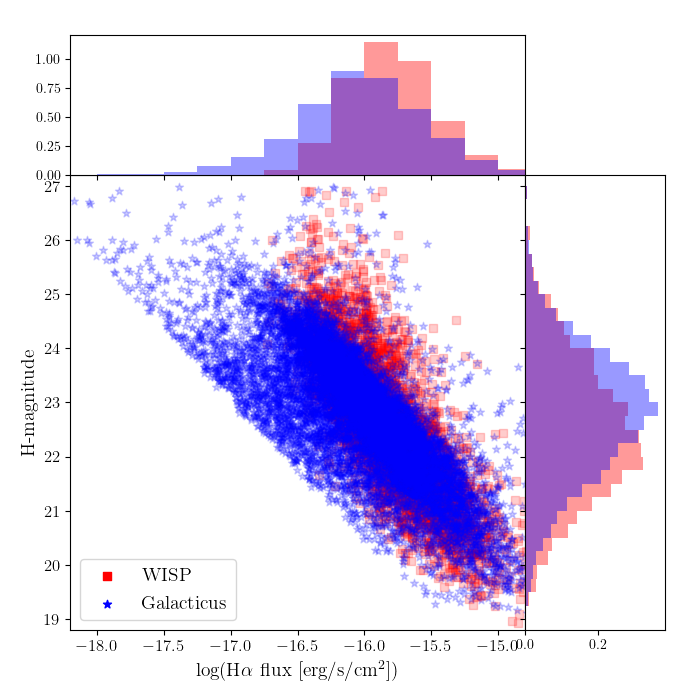}
\includegraphics[width=8cm]{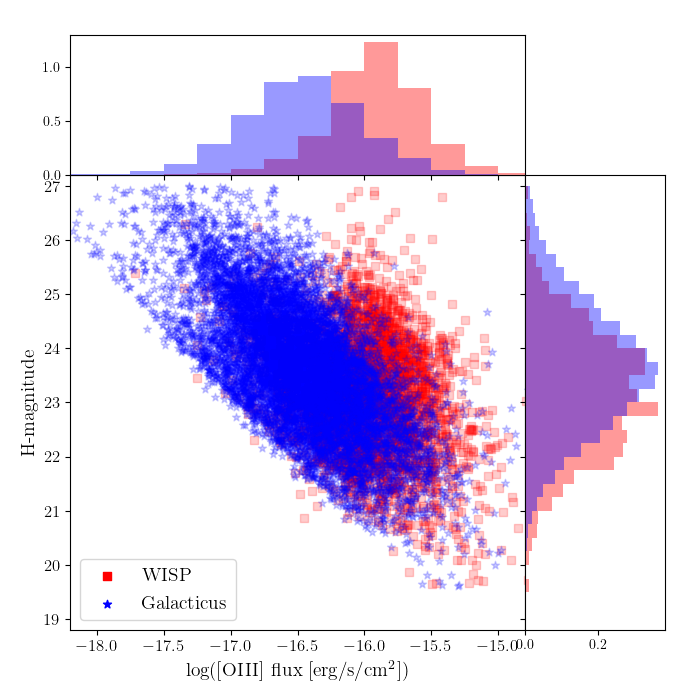}
\caption{Scatter plot for H$\alpha$ (left) and [OIII] (right) line flux and $H$-band magnitude, with the 1D projection.}
\label{fig:scatter}
\end{center}
\end{figure*}

\begin{figure*}
\begin{center}
\includegraphics[width=16cm]{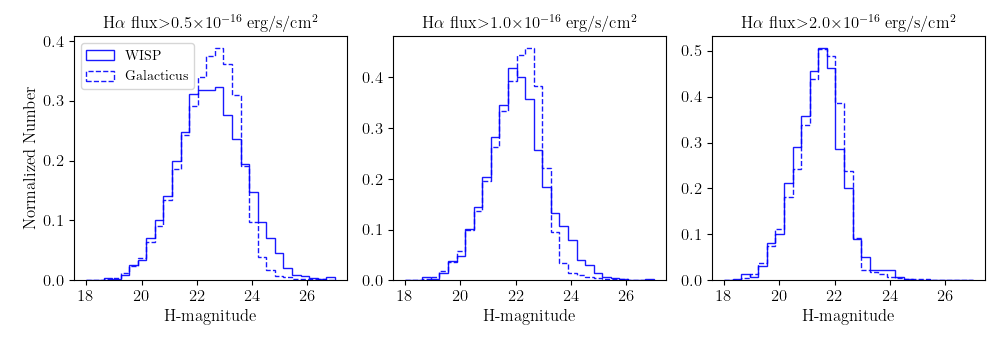}
\includegraphics[width=16cm]{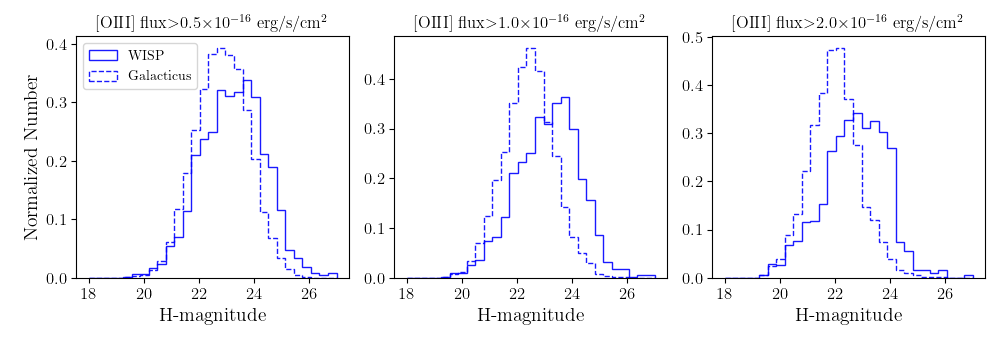}
\caption{Distribution of the $H$-band magnitude split by emission line flux: H$\alpha$ (top) and [OIII] (bottom)}
\label{fig:h_band_by_flux}
\end{center}
\end{figure*}

\begin{figure*}
\begin{center}
\includegraphics[width=16cm]{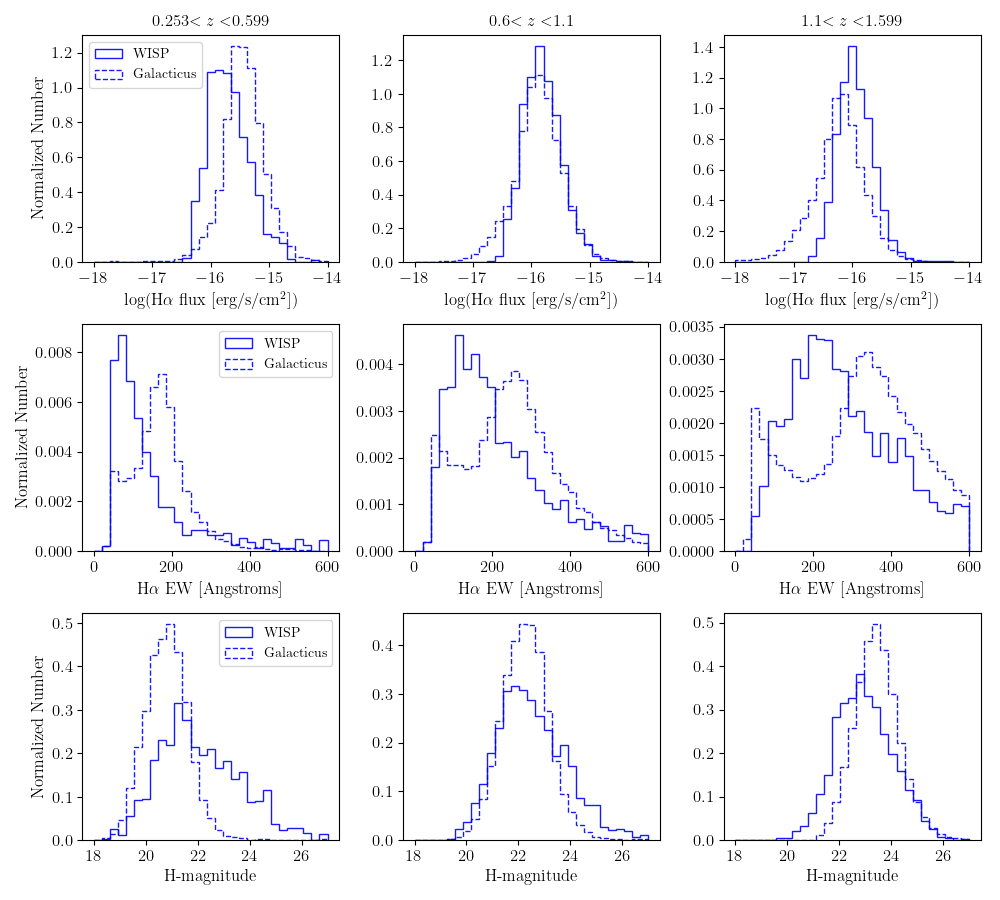}
\caption{Comparison of the H$\alpha$ emitting galaxies from WISP catalog and Galacticus prediction. The top panel is for line flux, the middle panel is for equilavent width, and the bottom panel is for $H$-band magnitude.}
\label{fig:Halpha_flux_EW}
\end{center}
\end{figure*}

\begin{figure*}
\begin{center}
\includegraphics[width=16cm]{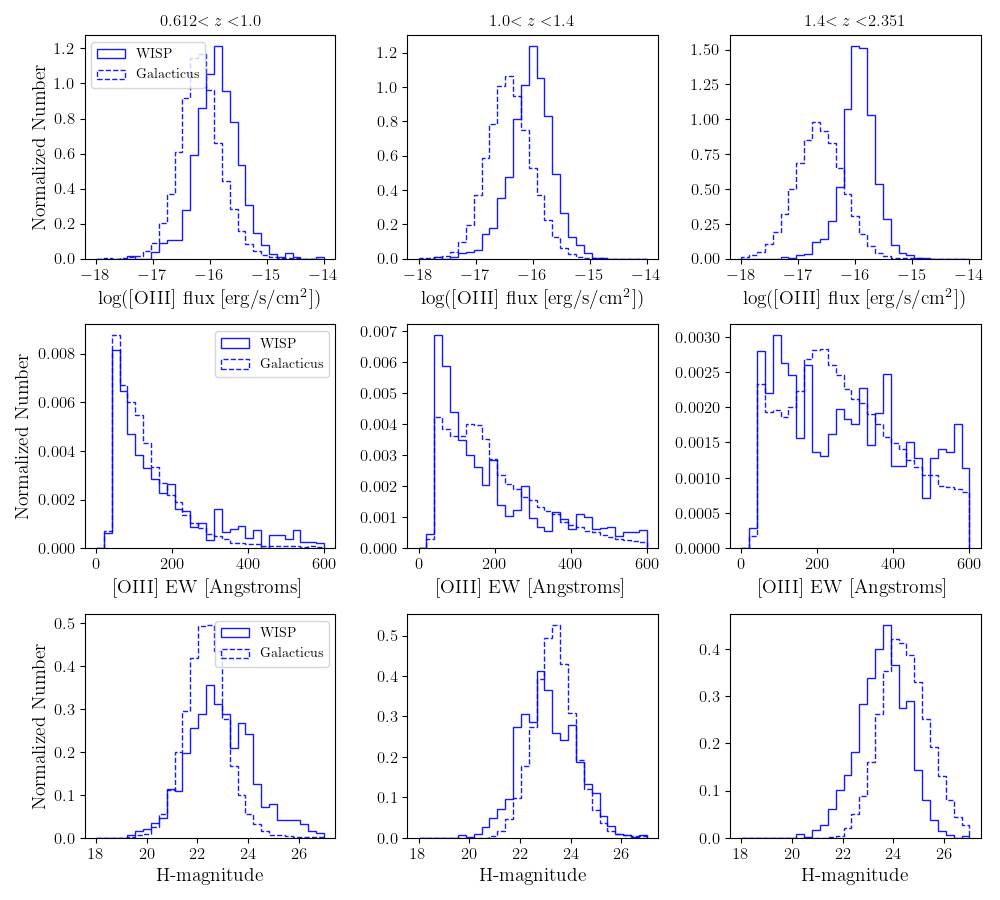}
\caption{The same as Figure \ref{fig:Halpha_flux_EW} but for [OIII].}
\label{fig:OIII_flux_EW}
\end{center}
\end{figure*}

Figure \ref{fig:WISP_completeness} shows that the completeness of the WISP galaxies can vary with multiple parameters.  For instance objects with H$\alpha$ EW lower than 30\AA~ are averagely ~20\% less complete than the high EW counterparts. The data point with error bars in the figure are obtained by splitting the sample into smaller bins to get the mean and standard deviation for each bin. In order to avoid systematics in the comparison, we apply the following cuts in the comparison: EW > 40\AA, the major axis < 0.8$^{\prime\prime}$. In order to make a fair comparison, we randomly downsample the Galacticus mock galaxies to the redshift distribution of WISP (Figure \ref{fig:z_distribution}) such that the two catalogs have the same redshift distribution for H$\alpha$ and [OIII] separately.

For the Galacticus mock, we first define the galaxy radius as
\begin{equation}
\text{max}(R_{\text{disk}}, R_{\text{spheroid}}),
\end{equation}
the maximum of the radius of the two components: disk and spheroid of the model galaxy. Then we assume a Planck 2016 cosmology to get the angular diameter distance of each objects and convert the galaxy size to angular scale so that the same cut on major axis < 0.8$^{\prime\prime}$ can also be applied. Note that this definition of galaxy size can be biased high since it always chooses the higher value. Another option can use stellar mass or luminosity: galaxy size equals the radius of the component (disk or spheroid) with higher stellar mass or luminosity. Based on the Galacticus mock, we find that this change can only affect up to 4\% of the galaxies. Therefore this doesn't impact the following comparison significantly. In order to avoid numerical issues in the mock catalog, we further exclude galaxies with stellar mass lower than $10^{7}~M_{\odot}$ and host halo mass lower than $10^{11.5}~h^{-1}M_{\odot}$. This mainly removes small galaxies with faint brightness and avoid issues of mass limit in the N-body simulation. 

In order to see the overall impact on the galaxy population from these cuts, we recalculate the H$\alpha$ luminosity function from the mock catalog, as shown in the bottom panel of Figure \ref{fig:Halpha_LF}. Note that the redshift downsampling is not  used here, in order to make comparisons with HiZELS. The results agree with our expectation that the dominant impact is on the faint galaxies. The overall amplitude of the luminosity function for bright galaxies remains the same. An interesting feature is that the impact can become more important for high redshift galaxies. For instance, the turn-over scale of the luminosity function increases from $\log{L_{H\alpha}[\text{erg s}^{-1}]}\sim41.5$ at z=0.4 to $\log{L_{H\alpha}[\text{erg s}^{-1}]}\sim42.3$ at z=2.23. Galaxies above this scale are rarely affected by the selections with the only exception of the lowest redshift measurement. 

We present the comparison of the $H$-band magnitude and emission line flux in Figure \ref{fig:scatter}. Note that the WISP catalog doesn't resolve H$\alpha$ and [NII], so our modeled H$\alpha$ is blended with [NII] using the empirical model of \cite{Faisst_2018}. In addition, WISP doesn't resolve the doublet of [OIII] either, and the catalog reports [OIII]$\lambda\lambda4959,5007$. Therefore our Galacticus prediction also combines contribution from these two lines. Not surprisingly, the results show that Galacticus correctly captures the overall correlation between the broad band and emission line luminosities. The brightness of mock galaxies is consistent with WISP observation within $1\sigma$. When projected into the 1D distribution of $H$-band magnitude and emission line flux, we find that the mock catalog is fainter than observations, but the dispersion is at a similar level of $< 1 \sigma$. 

In order to further examine the performance of the mock catalog, we split the comparison with different limits of emission line flux. Figure \ref{fig:h_band_by_flux} displays the distribution of $H$-band magnitude with different emission line limits. The results on the top panel for H$\alpha$+[NII] show excellent agreement between Galacticus and WISP regardless of the flux limit. This is as expected, since the line flux cut removes galaxies at the faint end of Figure \ref{fig:scatter} and thus brings the observed and Galacticus-predicted distributions closer. The overall consistency of [OIII] lines (bottom panel) is slightly worse than H$\alpha$+[NII] and the discrepancy can be more significant for brighter galaxies. This is also shown in the scatter plot of Figure \ref{fig:scatter} where the 1D distribution of [OIII] flux shows offset at the bright end. For bright [OIII] ELGs, our Galacticus model under-predicts the $H$-band magnitude by roughly 1~mag but this difference is within $1\sigma$. This discrepancy may be mitigated by the use of an improved dust model but that would also require some fine-turning of the model. 

In addition to the above cuts, we can also consider the continuum cut on the $H$-band magnitude: $H<25$ on both WISP and Galacticus catalog. Our test shows that this additional cut doesn't have significant impact on the overall comparisons.

Next, We split galaxies into three redshift subsamples: $0.253<z<0.6$, $0.6<1.1$ and $1.1<z<1.6$, and compare their distribution of H$\alpha$+[NII] emission line flux, EW and $H$-band magnitude in Figure \ref{fig:Halpha_flux_EW}. The top panel shows that the Galacticus model predicts less evolution of emission line flux with redshift than WISP measurements. This shows some mild offset in the low and high redshift subsamples, but the overall consistency still holds. The middle panel shows the distribution of EW. The cut on EW > $30\AA$ leads to a rapid drop at the faint end for both WISP and Galacticus catalog instead of an exponential-like behavior. The overall discrepancy is slightly larger than the emission line flux, indicating that Galacticus model underestimates the continuum level of the galaxies. Another noticeable feature is that Galacticus model shows slightly bi-modal distribution of EW at higher redshift, originating from the two branches of the galaxy population which is also shown in Figure \ref{fig:scatter}. This is likely related to the two sequences on the galaxy star formation rate and halo mass due to periods of mass loss, see discussion in \cite{Zhai_2021}. The bottom panels of Figure \ref{fig:Halpha_flux_EW} display the distribution for $H$-band magnitude. In contrast to the H$\alpha$ line flux in the top panel, Galacticus predicts slightly more evolution in the $H$-band distribution with redshift than the WISP observations, and the Galacticus model prediction is more concentrated. Overall, the Galacticus model is in reasonable agreement with the H$\alpha$ flux and $H$-band magnitude distributions, but performs less well at matching H$\alpha$ EW distributions, where it significantly overpredicts the typical EW. This suggests that possibly the H$\alpha$ emission is not arising the from ``correct'' galaxies in Galacticus---that is, strong H$\alpha$ emission is occurring in galaxies which are too low mass, resulting in a lower continuum and so an excessively high EW. Alternatively this discrepancy may highlight a limitation of our dust model, which currently applies the same extinction to both emission lines and continuum. Emission lines should likely experience greater extinction (since they arise from dense nebular regions) than the continuum, which would tend to reduce the predicted EWs. Overall, this discrepancy highlights the challenge of simultaneously modelling many different properties of the galaxy population.

Figure \ref{fig:OIII_flux_EW} shows the analogous analysis for [OIII]$\lambda\lambda4959,5007$. The galaxies are split into $0.612<z<1.0$, $1.0<z<1.4$ and $1.4<z<2.351$ subsamples. The aforementioned lower emission line flux from Galacticus is clearly shown in the top panel, and the offset is more significant at higher redshift. However, the distribution of EW shows prefect agreement between Galacticus and WISP, except for the high redshift subsample which is significantly affected by shot noise due to sample size. This indicates that our Galacticus model under-predicts the strength of [OIII] emission and continuum with a similar level. The distribution for $H$-band magnitude in the bottom panel is similar to the H$\alpha$ result in Figure \ref{fig:Halpha_flux_EW}. The deviations between the two catalogs exist, but they are within 1$\sigma$.

\subsection{Comparison with MOSDEF}

\begin{figure*}
\begin{center}
\includegraphics[width=8cm]{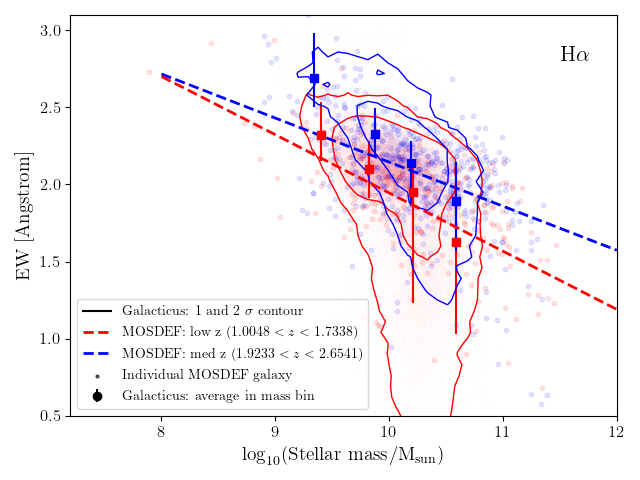}
\includegraphics[width=8cm]{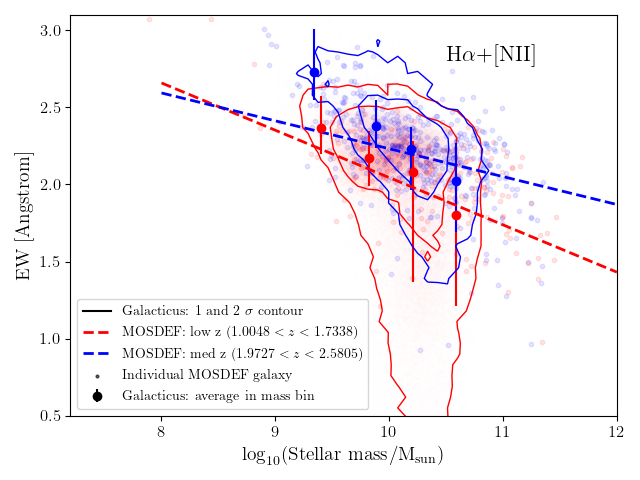}
\caption{Comparison of equivalent width for H$\alpha$ (left) and H$\alpha$+[NII] (right) with MOSDEF observations.}
\label{fig:EW_MOSDEF}
\end{center}
\end{figure*}

\begin{figure}
\begin{center}
\includegraphics[width=8cm]{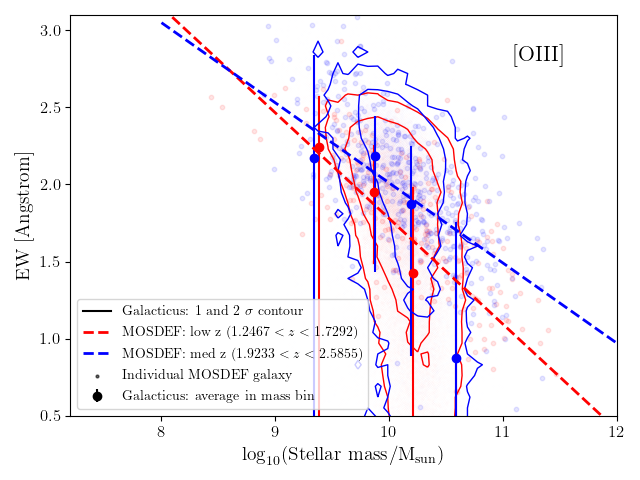}
\caption{Comparison of equivalent width for [OIII] with MOSDEF observations.}
\label{fig:EW_MOSDEF_OIII}
\end{center}
\end{figure}

The MOSDEF catalog measures EW for multiple emission lines, such as Balmer series, [OII], [OIII], [NII] lines. The derivation of stellar mass, age, color excesses and so on enables an exhaustive exploration of their correlation with emission lines. For simplicity, we choose the relation between EW and stellar mass to make the comparison with Galacticus.

The left hand panel of Figure \ref{fig:EW_MOSDEF} shows the distribution of H$\alpha$ EW and stellar mass. Since MOSDEF can resolve the contamination of [NII] to H$\alpha$, our Galacticus prediction is for H$\alpha$ only. We choose the galaxies at $z<3$ in this analysis to be consistent with the Roman redshift coverage, corresponding to the low-redshift and middle-redshift subsamples of MOSDEF data. We present the raw distribution of the MOSDEF galaxies as light dots in the background, and the dashed line denotes the best fit linear relation from \cite{Reddy_2018}. The Galacticus prediction is shown as contour plots covering 68\% and 95\% of the galaxies. Then we split galaxies by their stellar mass and obtain the mean and standard deviation of H$\alpha$ EW in each bin, as shown by the symbols with error bars. Clearly, MOSDEF data and Galacticus predictions are quite consistent. Galacticus prediction for the H$\alpha$ EW as a function of stellar mass matches the overall amplitude measured from MOSDEF; the agreement is similar to the the WISP comparison in previous section. The Galacticus prediction of the dependence of EW on the stellar mass agrees with observation, i.e., more massive galaxies have lower EW on average since the continuum flux is positively correlated with stellar mass but inversely proportional to EW. A slight offset from Galacticus prediction is the higher slope of the EW dependence on stellar mass measured by MOSDEF. The least massive galaxies prefer higher values of EW than predicted by Galacticus, and the tension is about $1\sigma$ or a bit more. This indicates that Galacticus predicts higher star formation rate for these less massive galaxies than seen in observations. Another noticeable feature of the Galacticus prediction is the redshift evolution, consistent with MOSDEF observations. At fixed stellar mass, galaxies at higher redshifts also have higher EW. This is associated with the redshift dependence of star formation rate at a given mass, see earlier investigations in \cite{Fumagalli_2012, Sobral_2014, Reddy_2018}.

In the right hand panel of Figure \ref{fig:EW_MOSDEF}, we present the result for blended H$\alpha$+[NII]. The overall consistency is similar to H$\alpha$, indicating that empirical correction of [NII] contamination is able to provide a good description given the current uncertainty of observations. 

As another primary emission line for the Roman HLSS, we present the comparison of EW for [OIII] line in Figure \ref{fig:EW_MOSDEF_OIII}. To be consistent with the MOSDEF analysis, the model prediction of [OIII] line combines the measurements of the doublet [OIII]$\lambda\lambda4959,5007$. Note that the redshift cuts are slightly different than H$\alpha$, but is consistent with the MOSDEF results (\citealt{Reddy_2018}). We can see that our model correctly captures the main characters of stellar mass dependence and redshift dependence of EW, but with mild offsets for the amplitude and slope. Due to the sample size and wide distribution, the Galacticus galaxies have larger dispersion and uncertainties than MOSDEF catalog. Compared with H$\alpha$, the [OIII] result has higher slope for the EW dependence on stellar mass. We should note that the continuum flux of the galaxy is tightly correlated with stellar mass, however the impact on different emission lines can vary. Therefore we do not expect Galacticus to give the same variation for all the emission lines considered, see a comprehensive discussion in \cite{Reddy_2018}.

Our analysis in this section shows that the Galacticus model can provide predictions for ELG H mag, H$\alpha$ and [OIII] line fluxes and EW, all consistent with observational data. It is also possible to investigate the dependence on other galaxy properties such as star formation rate, age, etc as in \cite{Reddy_2018}, which we leave to future work. 

\section{Photometric completeness of future surveys}

\begin{figure*}
\begin{center}
\includegraphics[width=8.0cm]{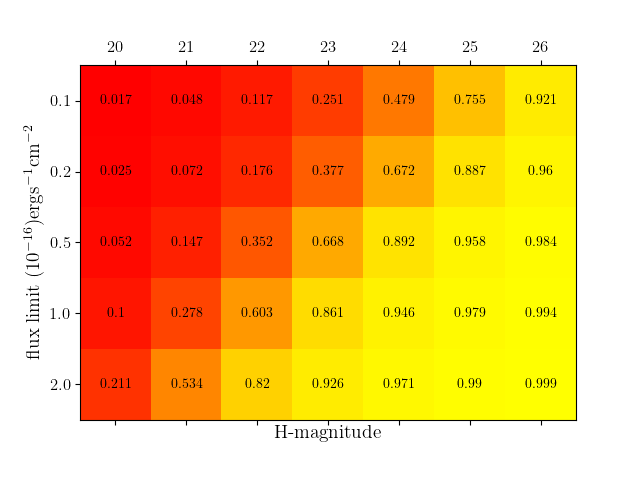}
\includegraphics[width=8.0cm]{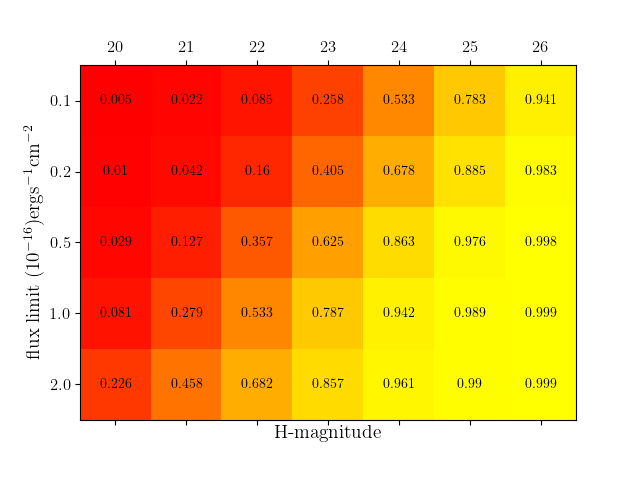}
\caption{Completeness of H$\alpha$ (left) and [OIII] (right) emitters from Galacticus mock. The color represents the completeness estimate, which is also shown as the numbers (no redshift downsampling based on WISP or other cuts). }
\label{fig:Hmag_cumu}
\end{center}
\end{figure*}

\begin{figure}
\begin{center}
\includegraphics[width=8cm]{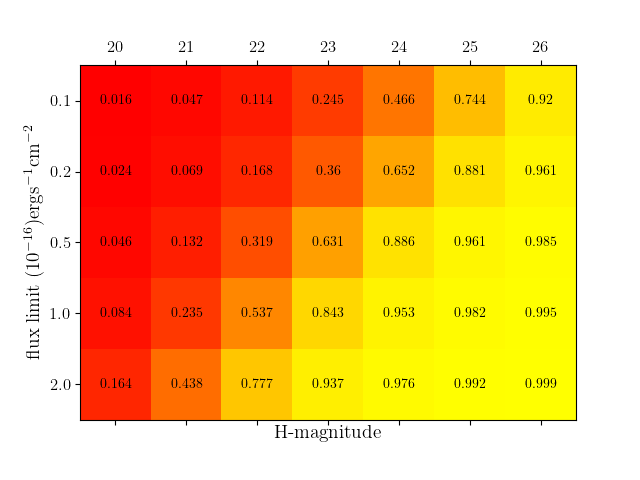}
\caption{Completeness of H$\alpha$+[NII] emitters from Galacticus.}
\label{fig:Hmag_cumu_NII}
\end{center}
\end{figure}

\begin{figure*}
\begin{center}
\includegraphics[width=7.0cm]{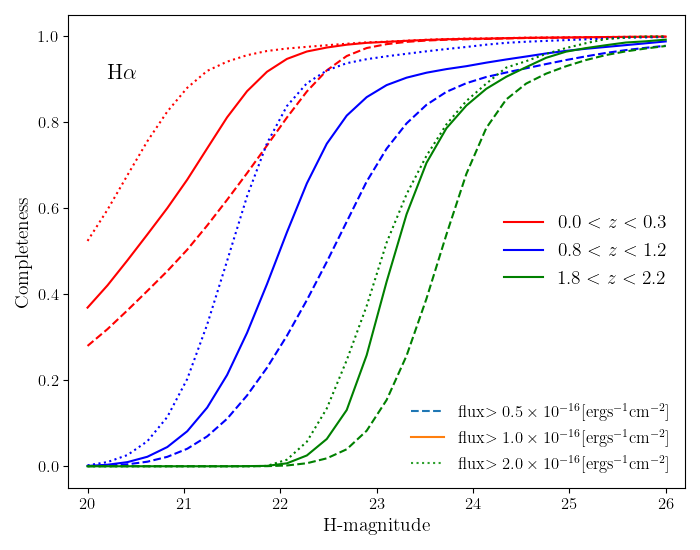}
\includegraphics[width=7.0cm]{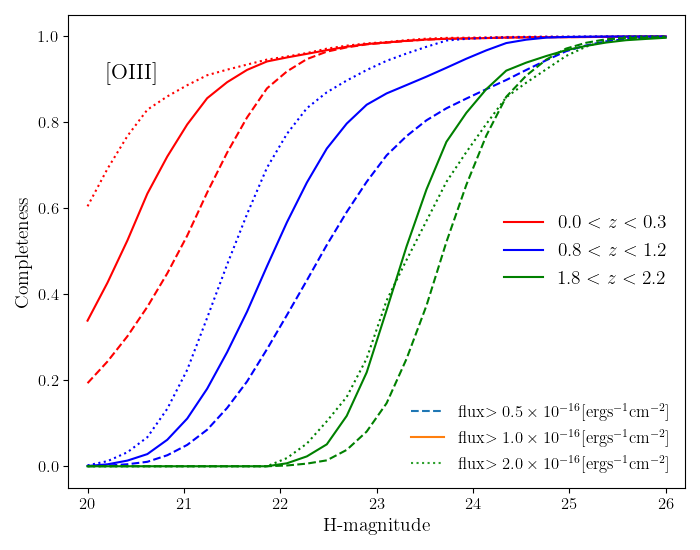}
\caption{Completeness of H$\alpha$ (left) and [OIII] (right) emitters from Galacticus as a function of $H$-band magnitude, for different redshift ranges. The colors represent redshift ranges as shown in the legend, while the line shapes denote flux limit.}
\label{fig:Hmag_cumu_z}
\end{center}
\end{figure*}

For slitless spectroscopic surveys like Roman HLSS, our Galacticus simulated galaxy catalog can provide important forecasts beyond the number density, clustering signal, cosmological implications and so on. It can also be used to estimate the photometric completeness given emission line and broad band luminosities, since the selection and spectral extraction of galaxies in a slitless spectroscopic survey is not only based on emission line strength, but also on the broad-band photometry used for constructing the decontamination model. 
The photometric completeness of a survey, as a function of the given line flux limit, provides key information for modeling observational systematics, as well as critical guidance in the data analysis. Note that the photometric completeness in this work is different from other similar completeness definitions, for instance the completeness defined in the WISP catalog which accounts for the loss of objects in the extraction process, the completeness for Roman HLSS galaxy clustering measurement which measures the fraction of galaxies with reliable redshift determinations.

We define the photometric completeness as the fraction of galaxies above an emission line threshold $f_{\text{limit}}$ for a given broadband magnitude $H_{\text{cut}}$, with respect to all galaxies with line flux above $f_{\text{limit}}$:
\begin{equation}
\text{photometric completeness} = \frac{N(H<H_{\text{cut}}, f>f_{\text{limit}}) }{N(f>f_{\text{limit}})},
\end{equation}
where $N$ denotes the galaxy number density, $H$ refers to $H$-band magnitude and $f$ is the emission line flux. This parameter can be expressed as a function of $H_{\text{cut}}$ and the emission line flux cut. By definition, this photometric completeness approaches 1 when the $H$-band cut is sufficiently faint, i.e. all galaxies above the emission line flux can be observed. An ideal survey has a high sensitivity to the emission line flux, and sufficiently faint broadband photometry. However, for slitless spectroscopic surveys, extracting sources too faint in photometry is counter-productive, as these are most prone to spectral contamination (e.g., spectral overlap) that can't be properly modeled. Thus there is a balance in choosing a sufficiently faint $H_{\text{cut}}$ such that the photometric completeness is high enough to enable good statistics for science, while avoiding making $H_{\text{cut}}$ so faint that noise dominates. Our work here is the key input for that evaluation.

In Figure \ref{fig:Hmag_cumu}, we present the forecast of photometric completeness for H$\alpha$ and [OIII]$\lambda5007$. The mock galaxies are trimmed within $0<z<3$ to match the upper limit of redshift for Roman HLSS. We can see that the dependence on $H$-magnitude cut is consistent with expectation. For a given limit on emission line flux, the galaxy sample becomes more complete for fainter $H$ mag cuts. For the nominal flux cut on H$\alpha$ emission line of Roman HLSS $f_{\mathrm{limit}}=1.0\times10^{-16}\mathrm{erg}~\mathrm{s}^{-1}\mathrm{cm}^{-2}$, the survey can be more than 94\% complete with $H_{\text{cut}}=24$ and approaches 100\% at $H_{\mathrm{cut}}=26$. For a much deeper survey with $f_{\mathrm{limit}}=0.5\times10^{-16}\mathrm{erg}~\mathrm{s}^{-1}\mathrm{cm}^{-2}$, the $H$-band magnitude needs to be at least 25 to reach 95\% completeness. For a very deep survey with $f_{\text{limit}}=1.0\times10^{-17}\mathrm{erg}~\mathrm{s}^{-1}\mathrm{cm}^{-2}$, the galaxy sample is only 50\% complete at $H_{\text{cut}}\sim24$. This can inevitably introduce unknown bias in the clustering analysis. With the constructed Euclid-like catalog from WISP, \cite{Bagley_2020} estimate that the completeness for H$\alpha$+[NII] emitters can reach 98\% at $H_{\text{cut}}=24$. In Figure \ref{fig:Hmag_cumu_NII}, we remeasure the completeness for the same line combination H$\alpha$+[NII]. We can see that this boost on the emission line flux can slightly change the completeness, and our prediction at $f_{\mathrm{limit}}=2.0\times10^{-16}\mathrm{erg}~\mathrm{s}^{-1}\mathrm{cm}^{-2}$ and $H_{\text{cut}}=24$ is in excellent agreement with their result.

The right hand panel of Figure \ref{fig:Hmag_cumu} shows the photometric completeness for [OIII] emission line. Compared with H$\alpha$, the completeness is lower on average. For Roman HLSS with $f_{\mathrm{limit}}=1.0\times10^{-16}\mathrm{erg}~\mathrm{s}^{-1}\mathrm{cm}^{-2}$, the [OIII] completeness is above 94\% for $H_{\text{cut}}=24$ and fainter. For an Euclid-like survey, the completeness can be higher than 96\%. 
 
 We further split the mock catalog by redshift ranges, and present the dependence of photometric completeness on redshift in Figure \ref{fig:Hmag_cumu_z}. Not surprisingly, lower redshift galaxies are more complete for a given emission line flux cut and $H$-band magnitude cut. For a deep H$\alpha$ ELG survey with $f_{\mathrm{limit}}=0.5\times10^{-16}\mathrm{erg}~\mathrm{s}^{-1}\mathrm{cm}^{-2}$, the nearby galaxies can be 80\% complete with $H_{\text{cut}}=22$, while galaxies at $z\sim2$ need $H_{\text{cut}}=24$ to reach the same level of photometric completeness. The [OIII] ELGs show a similar level of variation across redshifts. For shallower surveys like Euclid, we see that $H_{\text{cut}}=24$ can guarantee a $\sim90\%$ completeness for H$\alpha$ galaxies over the entire redshift range.

\section{Discussion and Conclusion}

Future galaxy redshift surveys like Roman HLWASS and Euclid will use emission line galaxies to explore the large scale structure in the Universe. It is of critical importance to understand the properties of these galaxies and maximize the science return in the study of cosmic acceleration and dark energy. This requires realistic forecast for such galaxies to help optimize the survey strategies. This goal can be realized via available observational data or numerical simulation. With the former, we can construct galaxy catalog using available data that can mimic the future surveys in terms of redshift range, selection function and spectral resolution. With the latter, we can perform high-resolution simulation of galaxies that can have reasonable prediction for statistics calibrated with observational data. The resultant product can predict the number density of target galaxies and clustering property as a function of redshift and survey depth. On the other hand, these future surveys will experience a complicated process of selection function related to emission line contamination, signal-to-noise ratio, galaxy size, redshift error and so on. Therefore the realistic forecast for these surveys requires extensive tests on different galaxy properties.

In this paper, we present a galaxy mock created using the Galacticus SAM and explore the ELG properties. This model was calibrated to match the available observed luminosity function of emission line galaxies. It can then make predictions of the galaxy properties that future surveys can expect to measure. With this galaxy mock, we first compare the emission lines and continuum magnitude with current observational data from WISP and MOSDEF surveys. This includes the emission line flux, equivalent width, stellar mass and $H$-band magnitude with appropriate cuts on the galaxy properties. This guarantees that the mock is as close to the observational sample as possible. The results for H$\alpha$ and [OIII] emission lines are quite consistent with observations. We also notice some offsets in the comparison, but the statistical discrepancy is within $1\sigma$. Thus our Galacticus galaxy mock has been validated by observational data. 

Our Galacticus model has predictive power. It captures the dependence of galaxy properties on redshift, e.g. the EW-stellar mass relation changes in the same way as MOSDEF observations, implying the robustness of the modeled star formation process over a wide redshift range.

Among the model parameters, dust attenuation has a direct impact on our results. We choose the \cite{Calzetti_2000} dust attenuation model with parameters determined in the SAM calibration process (\citealt{Zhai_2019MNRAS}), which can give consistent prediction of H$\alpha$ number counts compared with WISP (\citealt{Atek_2010}). This model applies to both the emission line and continuum. Therefore the predicted equivalent widths of the emission line are indeed dust-free, since the dust attenuation on the emission line and continuum cancels out at the same wavelength. We note that this is a simplified assumption, see some observational effort, e.g. \cite{Reddy_2015}. Since the emission lines and continuum can come from different components of galaxies with different star forming activities, it is possible that the emission lines and continuum can have different strengths of dust attenuation. Earlier attempt shows that the dust extinction is higher in the nebular regions than the general star population (\citealt{Calzetti_2000}). Incorporating this factor into our model can lower the EW prediction and thus increase the agreement with observations, see the middle rows of Figure \ref{fig:Halpha_flux_EW} and \ref{fig:OIII_flux_EW}. On the other hand, taking this effect into account can introduce additional degrees of freedom to our model, but our current comparison can serve as a preliminary examination on this assumption.

Based on this simulated catalog, we have forecasted the photometric completeness of H$\alpha$ and [OIII] emission line galaxies as a function of line flux limit and $H$-magnitdue cut. Our model predicts that the H$\alpha$ ELG can be $>$97\% complete with $H_{\text{cut}}=24$ for Euclid-like surveys, in agreement with the WISP based work by \citealt{Bagley_2020}.
For Roman HLSS, the nominal sensitivity with line flux limit $f_{\mathrm{limit}}=1.0\times10^{-16}\mathrm{erg}~\mathrm{s}^{-1}\mathrm{cm}^{-2}$ can reach a completeness of 94\% and above for both H$\alpha$ and [OIII] emitters at $H_{\text{cut}}=24$. A fainter limit of $H_{\text{cut}}=25$ can increase the completeness to 98\% and above but also means that the faint galaxies in photometry can be noise-dominated and thus degrade the galaxy statistics. Our result of [OIII] galaxies is similar to H$\alpha$ in terms of the completeness prediction and the dependence on line flux limit and $H$-magnitude cut. Note that however, [OIII] line is generally fainter than H$\alpha$ over the same redshift range.

The galaxy mock we have created can be used as the input catalog for spectroscopic simulations for Roman, Euclid, etc. The Galacticus model can trace star formation history of the galaxies and output their SEDs in a given wavelength range for both emission lines and continuum. This enables the construction of broad band photometries of various surveys to further test the model performance. In addition, we can use these SEDs and other physical properties to produce grism simulation for Roman HLWASS including both the direct slitless images and spectrum. The following extraction of individual object from this simulation can evaluate the survey performance under different conditions of roll angles and dithering patterns. The comparison with the input catalog and analysis of the redshift determination can tell us the fraction of galaxies that can contribute to the final clustering sample, which will be important for modeling the observed galaxy clustering to probe dark energy.

% End of mnras_template.tex

\section*{Data Availability}
The original dark matter halo catalogs are available from the UNIT simulation website. The galaxy mocks are available by request. A public webpage presenting the Galacticus mocks will be available at a later time.

\section*{Acknowledgements}

This work is supported in part by NASA grant 15-WFIRST15-0008, Cosmology with the High Latitude Survey Roman Science Investigation Team (SIT). ZZ would like to thank Anahita Alavi, Andreas Faisst and Xin Wang for helpful and valuable discussions in the observational analysis of  emission line galaxies, and Matt Malkan for his helpful comments on a draft of this paper. We are grateful to the MOSDEF team for making their data available to us. The UNIT simulations have been done in the MareNostrum Supercomputer at the Barcelona Supercomputing Center (Spain) thanks to the  cpu time awarded by PRACE under project grant number 2016163937. This work used the Extreme Science and Engineering Discovery Environment (XSEDE), which is supported by National Science Foundation grant number ACI-1548562 (\citealt{XSEDE_2014})

\rm{Software:} Python,
Matplotlib \citep{matplotlib},
NumPy \citep{numpy},
SciPy \citep{scipy}.

\appendix

\bibliographystyle{mnras}
\bibliography{emu_gc_bib,software}

\bsp	% typesetting comment
\label{lastpage}
\end{document}